\documentstyle[erice99,12pt,graphicx]{article}

\newcommand{\be}[1]{\begin{equation} \label{(#1)}}
\newcommand{\ee}{\end{equation}}
\newcommand{\ba}[1]{\begin{eqnarray} \label{(#1)}}
\newcommand{\ea}{\end{eqnarray}}

\begin{document}
\title{Photoproduction of meson and baryon resonances in a chiral unitary
approach} 
\author{E. Oset$^{1}$, E. Marco$^{2}$, J.C. Nacher$^{1}$, J.A. Oller$^{3}$, 
J.R. Pel\'aez$^{4}$, A. Ramos$^{5}$, H. Toki$^{6}$} 
\address{$^1$ Departamento de F\'{\i}sica Te\'orica and IFIC, Universidad de
Valencia, Burjassot(Valencia), Spain}
\vspace{-0.5cm}
\address{$^2$ Physik-Department, Technische Universit\"at M\"unchen,
D-85747 Garching, Germany}
\vspace{-0.5cm}
\address{$^3$ Forschungzentrum J\"ulich, IKP(Th), D-52425 J\"ulich,
Germany.}
\vspace{-0.5cm}
\address{$^4$ Departamento de F\'{\i}sica Te\'orica. Universidad Complutense de
Madrid, 28040 Madrid, Spain}
\vspace{-0.5cm}
\address{$^5$ Department d'Estructura i Constituents de la Mat\'eria,
Universitat de Barcelona, Barcelona, Spain}
\vspace{-0.5cm}
\address{$^6$ Research Center for Nuclear Physics (RCNP), Osaka University,
Ibaraki, Osaka 567-0047, Japan}

%
\maketitle

\abstracts{ By means of a coupled channel non-perturbative  unitary approach, 
 it is possible to extend
the strong constrains of Chiral Perturbation Theory to higher energies.
In particular, it is possible to reproduce the lowest lying resonances
in meson-meson scattering up to 1.2 GeV using
the parameters of the $O(p^2)$ and $O(p^4)$ Chiral Lagrangian. The meson baryon
sector can also be tackled along similar lines.
We report on an update of these results showing some examples of photon induced
reactions where the techniques have been recently applied.}

\section{Introduction}

Chiral Perturbation Theory ($\chi PT$) \cite{weinberg} has proved 
very successful in order to
describe the physics of mesons at very low energies. The key point
of the whole approach is to identify the lightest pseudoscalar
mesons $\pi , K$ and $\eta$ 
as the Goldstone bosons associated to the chiral symmetry
breaking. These particles will be the only degrees of 
freedom at low energies and their interactions can be described
in terms of the most general effective Lagrangian which
respects the chiral symmetry constraints. 

So far as this is a 
low energy approach, the amplitude of a given process
is basically given as an expansion in
the external momenta over the scale of symmetry breaking 
$4\pi f_\pi\simeq 1.2\,$GeV.  The approach is known to provide
a good description of meson interactions up to about 500 MeV. 
However, if one is interested in resonances in particular, as it happens in 
meson spectroscopy, there is little that one can do with just plain $\chi PT$. 
The method that we expose here naturally leads to low lying
resonances and allows one to face many problems so far intractable within
$\chi PT$.

The method incorporates the following elements: 1) Unitarity is implemented exactly;
2) It can deal with coupled channels allowed with pairs of particles from the 
lightest octets of 
pseudoscalar mesons and ($\frac{1}{2}^+$) baryons; 3) A chiral
expansion in powers of the external four-momentum of the lightest 
pseudoscalars
is done for
Re $T^{-1}$, instead of the $T$ matrix itself as it is done in standard $\chi PT$.

We sketch here the steps involved in this expansion for the meson meson
interaction. One starts from a $K$ matrix approach in coupled channels where
unitarity is automatically fulfilled and writes 
\begin{equation}
T^{-1} = K^{-1} - i\,\sigma ,
\end{equation} 
where $T$ is the scattering matrix, $K$ is a real matrix in the physical 
region
and $\sigma$
is a diagonal matrix which measures the phase-space available for the intermediate
states
\begin{equation} 
\sigma_{nn}(s) = - \frac{k_n}{8\pi\sqrt{s}}\,\theta\left(s - (m_{1n} + m_{2n})^2\right),
\end{equation}
where $k_n$ is the on shell CM momentum of the meson in the intermediate state
$n$ and $m_{1n}$, $m_{2n}$ are the masses of the two mesons in the state $n$. The meson
 meson states considered here are $K\bar{K}$, $\pi\pi$, $\pi\eta$, $\eta\eta$,
 $\pi K$, $\pi\bar{K}$, $\eta K$, $\eta\bar{K}$. Since $K$ is real in the
 physical region, from eq. (1)
 one sees that $K^{-1}$ = Re $T^{-1}$. In non-relativistic
 Quantum Mechanics, in the scattering
 of a particle from a potential, it is possible to expand
 $K^{-1}$ in powers of the momentum
 of the particle at low energies as follows (in the s-wave for simplicity)
\begin{equation} 
\hbox{Re}\,T^{-1}\equiv K^{-1} = \sigma\cdot ctg\delta\, \propto -\frac{1}{a} 
+ \frac{1}{2}r_0 k^2 ,
\end{equation}
with $k$ the particle momentum, $a$ the scattering length and $r_0$ the effective
range.

The ordinary  $\chi$PT expansion up to $O(p^4)$ is given by \cite{weinberg}
\begin{equation}
T = T_2 + T_4 ,
\end{equation}
where $T_2$ is obtained from the lowest order chiral
Lagrangian, $L^{(2)}$, and is of $O(p^2)$, whereas $T_4$ contains one loop diagrams in the s, t, u
channels, constructed from the lowest order Lagrangian, tadpoles and the
finite contribution from the 
tree level diagrams of the $L^{(4)}$ Lagrangian and is $O(p^4)$. This last contribution, 
after a suitable renormalization, is just a polynomial, $T^{(p)}$.
Our $T^{-1}$ matrix, starting from eq. (4) is given by
\begin{eqnarray}
\label{t-1}
T^{- 1} &=& \left[T_2+T_4+...\right]^{-1}=
T_2^{- 1} [1 + T_4  T_2^{- 1}+...]^{- 1}\nonumber \\ 
&=& T_2^{- 1} [1 - T_4  T_2^{- 1}+...]=T_2^{-1} [T_2-T_4]
T_2^{-1}
\end{eqnarray}
Due to the fact that $\hbox{Im}\,T_4= T_2 \sigma T_2$, the above equation
is nothing but eq. (1), but using eq. (4) to expand  $K^{-1}=\hbox{Re}\, T^{-1}$.
Inverting the former result, one obtains:
\begin{equation}
T = T_2\,[ T_2 - T_4]^{-1}\, T_2,
\end{equation}
which is the coupled channel generalization of the inverse amplitude method of
\cite{dob}.

        Once this point is reached one has several options to proceed:

a) A full calculation of $T_4$ within the same renormalization scheme as in
$\chi PT$ can be done. The eight $L_i$ coefficients from $L^{(4)}$ are then fitted
to the existing meson meson data on phase shifts and inelasticities up to 1.2 GeV, where
4 meson states are still unimportant. This procedure has been carried out in
\cite{dob,gue}. The resulting $L_i$ parameters are compatible with those used in $\chi PT$. 
At low energies the $O(p^4)$ expansion for $T$ of eq. (6) is identical to that
in $\chi PT$. However, at higher energies the nonperturbative structure of 
eq. (6),
 which implements unitarity exactly, allows one to extend the information
 contained in the chiral Lagrangians to much higher energy than in ordinary 
$\chi$ PT. Indeed it
reproduces the resonances present in the L = 0, 1 partial waves.

\vskip .2cm

b) A technically simpler and equally successful additional approximation
 is generated by ignoring the crossed channel loops and
tadpoles and reabsorbing them in the $L_i$
coefficients given the weak structure of these terms in the physical region.
The fit to the data with the new $\hat{L}_i$ coefficients reproduces the whole meson
meson sector, with the position, widths and partial decay widths of the
$f_0(980)$, $a_0(980)$, $\kappa(900)$, $\rho(770)$, $K^\ast(900)$ resonances in good
agreement with experiment \cite{oller1}. A cut off regularization is used in \cite{oller1} for the
loops in the s-channel. By taking the loop function with two intermediate
mesons
\begin{equation}
G_{nn}(s) = i\int\frac{d^4 q}{(2\pi)^4}\, \frac{1}{q^2 - m_{1n}^2 + i\epsilon}
\, \frac{1}{(P-q)^2 - m_{2n}^2 + i\epsilon},
\end{equation}
where $P$ is the total meson meson momentum, one immediately notices that
\begin{equation}
\hbox{Im}\, G_{nn}(s) = \sigma_{nn}.
\end{equation}
Hence, we can write
\begin{equation}
\hbox{Re}\, T_4 = T_2\, \hbox{Re}\, G\, T_2 + T_4^{(p)},
\end{equation}
where $\hbox{Re}\, G$ depends on the cut off chosen for $|\vec{q}|$. This means that the
$\hat{L}_i$ coefficients of $T_4^{(p)}$ depend on the cut off choice, much as the
$L_i$ coefficients in $\chi PT$ depend upon the regularization scale.

\vskip .2cm

c) For the L = 0 sector (also in L = 0, S = $-1$ in the meson baryon interaction)
a further technical simplification is possible. In these cases it is possible
to choose  the cut off such that, given the relation between $\hbox{Re}\, G$
and $T_4^{(p)}$, this latter term is very well approximated by 
$\hbox{Re} T_4= T_2 \,\hbox{Re}\, G\, T_2$. This is possible in those cases
because of the predominant role played by the unitarization of the lowest
order $\chi PT$ amplitude, which by itself leads to the low lying resonances,
 and because other genuine QCD resonances appear at higher energies.

 In such a case eq. (5) becomes 
\begin{equation}
T = T_2 \,[T_2 - T_2\, G \,T_2]^{-1} \,T_2 = [1 - T_2 \,G]^{-1}\, T_2,
\end{equation}
or, equivalently,
\begin{equation}
T = T_2   + T_2 \,G \,T,
\end{equation}
which is a Bethe-Salpeter equation with $T_2$ and $T$ factorized on shell outside
the loop integral, with $T_2$ playing the role of the potential. This option has
proved to be successful in the L = 0 meson meson sector in \cite{oller2} and in the
L = 0, S = $-1$ meson baryon sector in \cite{osetra}.

        In the meson baryon sector with S = 0, given the disparity of the
        masses in the coupled channels $\pi N$, $\eta N$, $K\Sigma$,
        $K\Lambda$,
        the simple ``one cut off approach'' is not possible. In \cite{kaiser} higher
        order Lagrangians are introduced while in \cite{par} different subtraction
        constants in G are incorporated
        in each of the former channels leading in both cases to acceptable
        solutions when compared with the data.

In fig. 1 we show the results done with the method of \cite{oller1} for some selected phase shifts and inelasticities 
in the meson meson sector, showing resonances in different channels, (see \cite{phipipi} for an update of the results). The
agreement with the meson meson data is quite good up to 1.2 GeV and the
parameters $\hat{L_i}$ obtained from the fit are essentially compatible with
those of $\chi PT$.
\begin{figure}[ht]
\vspace{-.3cm}
\centerline{
\includegraphics[width=0.7\textwidth,angle=0]{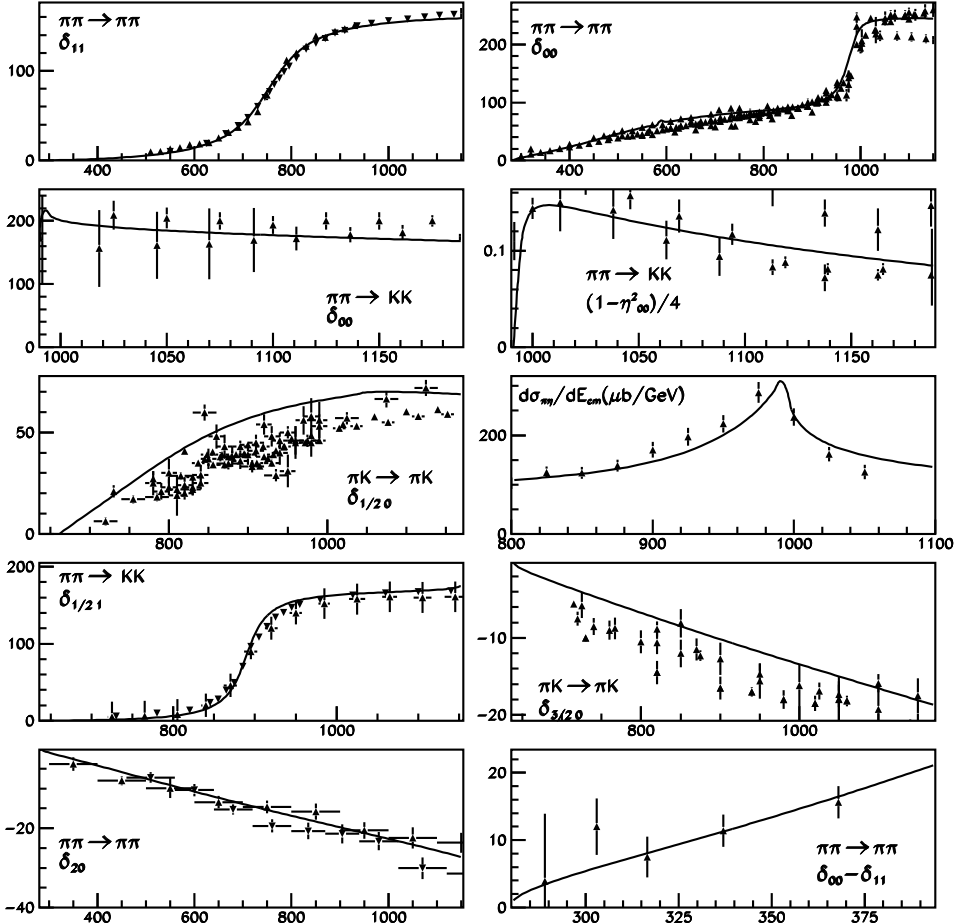}
}
\caption{ Meson-meson scattering results of the non-perturbative chiral 
approach.
For the data see references in \cite{dob,oller1,gue,ollernew}. Horizontal 
scale in MeV.
}
\label{fig3}
\vspace*{-.3cm}
\end{figure}

\section{$\bar{K}N$ interaction in free space}

The meson-baryon interaction Lagrangian at lowest order in momentum is given by
\begin{equation}
L_1^{(B)} = \langle \bar{B} i \gamma^{\mu} \frac{1}{4 f^2}
[(\Phi \partial_{\mu} \Phi - \partial_{\mu} \Phi \Phi) B
- B (\Phi \partial_{\mu} \Phi - \partial_{\mu} \Phi \Phi)]
\rangle     \ ,
\end{equation}
where $\Phi$ represents the octet of pseudoscalar mesons and $B$ the 
octet of $1/2^+$ baryons. The symbol $\langle \rangle$ denotes the trace of
SU(3) matrices.

The coupled channel formalism requires to evaluate the transition
amplitudes between the different meson-baryon channels.
For $K^- p$ scattering there are ten channels, namely $K^-p$, $\bar{K}^0 n$,
$\pi^0
\Lambda$, $\pi^0 \Sigma^0$,
$\pi^+ \Sigma^-$, $\pi^- \Sigma^+$, $\eta \Lambda$, $\eta
\Sigma^0$,
$K^+ \Xi^-$ and $K^0 \Xi^0$, while in the case of $K^- n$ scattering
there are six: $K^-n$, $\pi^0\Sigma^-$,
 $\pi^- \Sigma^0$, $\pi^- \Lambda$, $\eta
\Sigma^-$ and
$K^0 \Xi^-$. These amplitudes have the form
\begin{equation}
V_{ij}=-C_{ij}\frac{1}{4f^2}\overline{u}(p_i)\gamma^\mu u(p_j)
(k_{i\mu}+k_{j\mu}) \ ,
\end{equation}
where $p_j,p_i(k_j,k_i)$ are the initial, final momenta of the baryons
(mesons) and
$C_{ij}$ are SU(3) coefficients that can be found
in Ref.~\cite{osetra}.
At low energies the spatial components can be neglected and the
amplitudes reduce to
\begin{equation}
V_{i j} = - C_{i j} \frac{1}{4 f^2} (k_j^0 + k_i^0) \ .
\end{equation}

The coupled-channel BS
equations in the center of mass frame read
\begin{equation}
T_{i j} = V_{i j} + \overline{V_{i l} \; G_l \; T_{l j}} \ ,
\end{equation}
\noindent
where the indices $i,l,j$ run over all possible channels and
$\overline{V_{i l} \; G_l \; T_{l j}}$ corresponds to the loop integral
involving $V$, $T$ and the meson baryon propagators of $G$, all functions of the
loop variable.
However, as was shown in Ref.~\cite{osetra}, the off-shell part of $V_{il}$ and
$T_{lj}$
goes into
renormalization of coupling constants and $V_{il}$,
$T_{lj}$
factorize outside the integral with their on-shell values, thus reducing
the problem to one of  inverting
a set of algebraic equations. 

\begin{table}[ht]
\caption{$K^-p$ threshold ratios and $K^-N$ scattering lengths}
\begin{center}
\begin{tabular}{|l|c|c|}
\hline
 & This work & Exp.  \\ 
 \hline
\rule[-6mm]{0mm}{15mm}
$\gamma=\displaystyle\frac{\Gamma(K^-p\to \pi^+ \Sigma^-)}{
\Gamma(K^-p \to \pi^-\Sigma^+)}$ & 2.32 & 2.36$\pm$0.04
\cite{To71} \\
\hline 
\rule[-6mm]{0mm}{15mm}
$R_c=\displaystyle\frac{\Gamma(K^-p\to {\rm charged})}{
\Gamma(K^-p \to {\rm all})}$ & 0.627 & 0.664$\pm$0.011
\cite{To71} \\
\hline 
\rule[-6mm]{0mm}{15mm}
$R_n=\displaystyle\frac{\Gamma(K^-p\to \pi^0\Lambda)}{
\Gamma(K^-p \to {\rm neutral})}$ & 0.213 & 0.189$\pm$0.015 
\cite{To71} \\
\hline 
$a_{K^-p}$ (fm) & $-$1.00 + i 0.94 & $-$0.67 + i 0.64 \cite{Adm81}
\\
                &                & $-$0.98 (from Re($a$))
\cite{Adm81} \\
         &          & ($-$0.78$\pm$0.18) + i(0.49$\pm$0.37)
\cite{Iw97}\\
 & & \\
 \hline
$a_{K^-n}$ (fm) & 0.53 + i 0.62 & 0.37 + i 0.60 \cite{Adm81} \\
                &                & 0.54 (from Re($a$))\cite{Adm81} \\\hline
\end{tabular}
\end{center}
\label{tab:table1}
\end{table}

The value of the cut-off, 
$q_{\rm max}=630$ MeV, was
chosen to reproduce the $K^- p$ scattering branching ratios
at threshold, while the weak decay constant, $f=1.15
f_\pi$, was taken in between the pion and kaon ones to optimize
the position of the $\Lambda(1405)$ resonance. The predictions of the model for several
scattering observables are summarized in Table \ref{tab:table1}. Cross sections
for $K^- p$ scattering to different channels are also calculated in
\cite{osetra} and good results are obtained for low energies of the kaons where
the s-wave is dominant. 

 A recent application of these methods in the S=0 sector is done in
 \cite{ramonet}, where the $\Delta(1232)$ resonance is also nicely reproduced.

\section{Application to the photoproduction of meson baryon pairs in resonant
states}
As quoted above, a good description of the interaction of $K^-p$ and its coupled 
channels 
is obtained in terms of the lowest order Lagrangians and the Bethe Salpeter
equation with a single cut off. One of the interesting features of the approach
is the dynamical generation of the $\Lambda(1405)$ resonance just below the
$K^-p$ threshold. The threshold behavior of the $K^-p$ amplitude is thus
very much tied to the properties of this resonance. Modifications of these
properties in a nuclear medium can substantially alter the $K^-p$ and $K^-$
nucleus interaction and experiments looking for these properties are most welcome. Some 
electromagnetic reactions appear well suited for these studies.
Application of the chiral unitary approach to the
$K^-p\rightarrow\gamma\Lambda$, $\gamma\Sigma^0$ reactions at threshold has
been carried out in \cite{lee} and a fair agreement with experiment is found. In
particular one sees there that the coupled channels are essential to get a good
description of the data, increasing the $K^-p\rightarrow\gamma\Sigma^0$ rate
by about a factor 16 with respect to the Born approximation.

        In a recent paper \cite{nac1} the $\gamma p\rightarrow
        K^+\Lambda(1405)$ reaction was proposed as a means to study the properties of the
        resonance, together with the $\gamma A\rightarrow
        K^+\Lambda(1405) A'$ reaction to see the modification of its properties
        in nuclei. The resonance $\Lambda(1405)$ is seen in its decay products
        in the $\pi\Sigma$ channel, but as shown in \cite{nac1} the sum of the cross
        sections for $\pi^0\Sigma^0$, $\pi^+\Sigma^-$, $\pi^-\Sigma^+$
        production has the shape of the resonance $\Lambda(1405)$ in the I = 0
        channel. Hence, the detection of the $K^+$ in the elementary reaction,
         looking at $d\sigma/dM_I$ ($M_I$ the invariant mass of the meson
         baryon system which can be induced from the $K^+$ momentum), is
         sufficient to get a clear $\Lambda(1045)$ signal. In nuclear targets
         Fermi motion blurs this simple procedure (just detecting the $K^+$), but the
         resonance properties can be reconstructed by observing the decay
         products in the $\pi\Sigma$ channel. In fig. 2 we show the cross
         sections predicted for the $\gamma p\rightarrow K^+ \Lambda(1405)$
         reaction looking at $K^+\pi^0\Sigma^0$, $K^+ all$ and $K^+ \Lambda(1405)$
         (alone). All of them have approximately the same shape and strength
         given
         the fact that the I = 1 contribution is rather small. In the figure the
         dashed dotted line indicates what one should expect to see in nuclei,
         just detecting the $K^+$, from the effect of Fermi motion.
\begin{figure}[ht]
\centerline{
\includegraphics[width=0.4\textwidth,angle=-90]{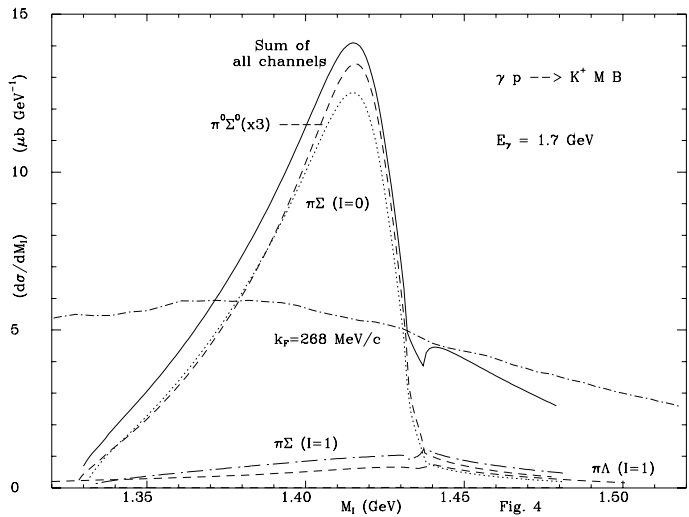}
}
\caption{Cross section for $\gamma p\rightarrow K^+ X$ with $X=all$,
$\pi^0\Sigma^0$, $\Lambda$(1405).
}
\label{fig3}
\end{figure}
         The energy chosen for the photon is $E_\gamma$ = 1.7 GeV which makes it
        suitable of experimentation at SPring8/RCNP, where the experiment is
        planned \cite{nakano}, and TJNAF.
        
        One variant of this reaction is the crossed channel reaction $K^- p
\rightarrow\Lambda(1405)\gamma$. This reaction, for a $K^-$ momentum
in the 300 to 500 MeV/c range, shows clearly the $\Lambda(1405)$ resonant production
\cite{nac2} and
has the advantage that the analogous reaction in nuclei still allows the
observation of the $\Lambda(1405)$ resonance with the mere detection of the
photon, the Fermi motion effects being far more moderate than in the case of
the $\gamma A\rightarrow K^+\Lambda(1405)X$ reaction which requires larger
photon momenta and induces a broad distribution of $M_I$ for a given $K^+$
 momentum.

\section{Photoproduction of resonant two meson states}

    Another application which can be done using the same reaction is the
 photoproduction of resonant two meson states. Particularly the $f_0$(980) and
 $a_0$(980) resonances. These states appear in $L=0$ in isospin zero and one
 respectively. The scalar sector of the meson is very controversial and the
 chiral unitary theory has brought a new perspective on these
 states. In particular, it has been possible to identify the lightest scalar
octet, made of the $\sigma$, $f_0(980)$, $a_0(980)$ and $\kappa(900)$ resonances.
{\em All of them can be simply generated by unitarization 
of the lowest order ChPT}, with just a cutoff as a free parameter.

The $O(p^4)$ chiral parameters
can be understood as the residual contact terms that appear
when  one integrates
out heavier states and the resulting Lagrangian is that of ChPT. Hence, the
values of the chiral constants can be related to the masses and widths
of the preexisting heavier resonances (``Resonance Saturation Hypothesis''). 
Indeed, most of their values
are saturated by vector resonances alone 
(that is vector meson dominance) but some other parameters still need the existence
of scalar states. 
Recently  \cite{ollernew}, using the N/D unitarization method with
explicit resonances added to the lowest order ChPT Lagrangian,  
it has been established that these heavier scalar
states should appear with a mass around 1.3 - 1.4 GeV for the octet and 1 GeV for the singlet.
In addition, the  $\sigma$, $\kappa$, $a_0$ and a strong contribution
to the $f_0$, were also generated from the unitarization of the ChPT lowest 
order.
These states still survive when the heavier scalars are removed.
That agrees with our observation that the $\sigma$, $\kappa$, $f_0$ and $a_0$
are generated {\em independently of the chiral parameters}, that is, of the preexisting 
scalar nonet, which is heavier. In addition, it was also stablished in that work that
the physical $f_0(980)$ resonance is a mixture between the discussed
strong $K\bar{K}$ scattering 
contribution and the preexisting singlet resonance with a mass around 1 GeV.
Since Chiral Perturbation Theory does not deal with quarks and
gluons, it is very hard to make any conclusive statement about the 
nature of these states
($q\bar{q}$, four-quark, molecule, etc...), unless we make additional assumptions. 
However, any model of the nature of these states should be able to explain the different
features of these resonances as they appear in the chiral unitary approach.

In addition, it would be very interesting to obtain further information from other processes.
In the present case the reaction suggested is \cite{marco}
 $\gamma p\rightarrow p M$.
where $M$ is either of the resonances $a_0$(980) or $f$(980). In practice the meson M
 will decay into two mesons , $\pi\pi$ or $K\bar{K}$ in the case of the 
 $f_0$(980)  or $K\bar{K}$, $\pi\eta$ in the case of the $a_0$(980). 
\begin{figure}[ht]
\centerline{
\includegraphics[width=0.4\textwidth,angle=-90]{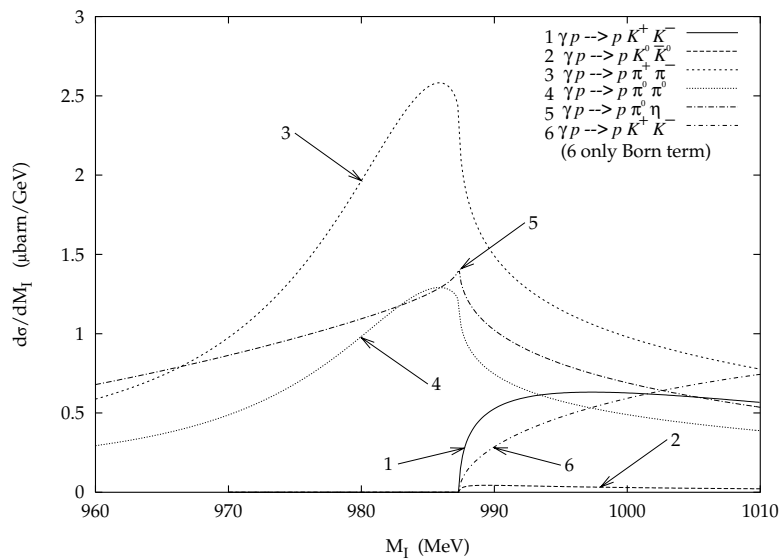}
}
\caption{Results for the photoproduction cross section on protons
as a function of the invariant
mass of the meson-meson system.
}
\label{fig3}
\end{figure}

In Fig.~3 we show the results for the 5 channels considered. We observe
clear peaks for $\pi^+ \pi^-$, $\pi^0 \pi^0$ and $\pi^0 \eta$
production around 980 MeV. The
peaks in $\pi^+ \pi^-$ and $\pi^0 \pi^0$ clearly correspond to the
formation of the $f_0(980)$ resonance, while the one in
$\pi^0 \eta$ corresponds to the formation of the
$a_0(980)$. The $\pi^0 \pi^0$ cross section is $\frac{1}{2}$
of the $\pi^+ \pi^-$ one due to the symmetry factor . 
The $K^+ K^-$ and $K^0 \bar{K}^0$ production cross section appears
at energies higher than that of the resonances and hence do not show
the resonance structure. Yet, final state interaction is very important
and increases appreciably the $K^+ K^-$ production cross section
for values close to threshold with respect to the Born approximation.

It is interesting to notice that the $f_0$(980) resonance shows up as a peak
in the reaction. This is in contrast to the cross section for
$\pi \pi \rightarrow \pi \pi$ in $I=0$ which exhibits a minimum at the
$f_0$ energy because of the interference between the $f_0$
contribution and the $\sigma(500)$ broad resonance. 

However, we should bear in mind that we
have plotted there the contribution of the $f_0$ resonance alone. The tree
level contact term and Bremsstrahlung diagrams, plus other contributions
which would produce a background, are not considered there. 

  In any case it is interesting to quote in this respect that a related reaction
  from the dynamical point of view, which also involves the interaction of two
  mesons in the final state, the $\phi \rightarrow \pi^0 \pi^0 \gamma$ 
  and the $\phi \rightarrow \pi^0\eta \gamma$ decay, which have been measured
  recently at Novosibirsk \cite{Novo},
   show clearly the $f_0$ and $a_0$ excitation,
  respectively, in the invariant mass spectra of the two mesons.
  A theoretical study along the lines reported here has been done in \cite{uge}
  where a good description of the experimental spectra as well as the absolute
  rates is obtained. 
\section{ Summary}

  We have reported on the unitary approach to meson meson and meson baryon
  interaction using chiral Lagrangians, which has proved to be an efficient
  method to extend the information contained in these Lagrangians to higher
  energies where $\chi PT$ cannot be used. This new approach has opened the
  doors to the investigation of many problems so far intractable with $\chi PT$
  and a few examples have been reported here. We have applied these techniques
  to the problem of photoproduction of scalar mesons $f_0$(980), $a_0$(980) and
  the photoproduction of the $\Lambda(1405)$, a resonant state of meson baryon
  in the $S=-1$ sector and have found signals which are well within measuring
  range in present facilities. The experimental implementation of these
  experiments confronted with the theoretical predictions will contribute with new
  tests of these emerging pictures implementing chiral symmetry and unitarity,
  which for the moment represent the most practical approach to QCD at low
  energies.

\section*{Acknowledgments.}
\vspace{-0.4cm}       
         This work is partly supported by DGICYT, contract
        number PB 96-0753.

\end{document}